
%
%
%
%
%
\input harvmac.tex
\def\hT{{\widehat T}}
\def\hH{{\widehat H}}
\def\hh{{\widehat h}}
\def\hA{{\widehat A}}
\def\hB{{\widehat B}}
\def\hv{{\widehat v}}
\def\tT{{\tilde T}}
\def\tH{{\tilde H}}
\def\th{{\tilde h}}
\def\tA{{\tilde A}}
\def\tB{{\tilde B}}
\def\tv{{\tilde v}}
\Title{PUPT--1284/ LAVAL-PHY-26/91}
{{\vbox {
\centerline{Singular Vectors and Conservation Laws}
\bigskip
\centerline{of Quantum KdV type equations}
}}}

\centerline{P. Di Francesco\foot{Work supported by NSF grant
PHY-8512793.}}
\bigskip\centerline
{\it Joseph Henry Laboratories,}
\centerline{\it Princeton University,}
\centerline{\it Princeton, NJ 08544.}
\bigskip
\centerline{and}
\bigskip
\centerline{P. Mathieu\foot{Work supported by NSERC (Canada) and
FCAR (Qu\'ebec).}}
\bigskip\centerline{\it D\'epartement de Physique,}
\centerline{\it Universit\'e Laval,}
\centerline{\it Qu\'ebec, Canada, G1K 7P4.}
\vskip .2in

\noindent
We give a direct proof of the relation between vacuum singular vectors
and conservation laws for the quantum KdV equation  or equivalently
for $\Phi_{(1,3)}$-perturbed conformal field theories.
For each degree at which a classical conservation law exists, we find
a quantum conserved quantity for a specific value of the central charge.
Various generalizations ($N=1,2$ supersymmetric, Boussinesq) of this
result are presented.

\Date{09/91}
%
\newsec{Introduction}

The equivalence between the Poisson Bracket associated with the second
Hamiltonian structure of the Korteveg-de Vries (KdV) equation, and the
classical form of the Virasoro algebra has the following implication:
there exists an infinite number of integrals in involution, whose
densities are differential polynomials of the classical energy-momentum
tensor
\ref\GER{J.L. Gervais, Phys. Lett.{\bf B160} (1985) 277.}.
The latter corresponds to the KdV field, and these integrals are the KdV
conservation laws. It is natural to conjecture that this result remains
valid in the quantum case
\ref\SAYA{R. Sasaki and I. Yamanaka, Adv. Stud. in Pure Math. {\bf 16}
(1988) 271.}.
This is equivalent to conjecture the complete integrability
of the quantum KdV (qKdV) equation formulated canonically via the
Virasoro algebra
\ref\KM{B.A. Kuperschmidt and P. Mathieu, Phys. Lett. {\bf B227} (1989)
245.}.
In the context of Conformal Field Theories, these integrals are trivially
conserved due to $\bar z$ independence of their densities.
However, for minimal models, it was conjectured
\ref\ZAMO{A.B. Zamolodchikov, JETP Lett. {\bf 46} (1987) 160;
Int. Jour. Mod. Phys. {\bf A3} (1988) 4235; Adv. Stud. in Pure Math.
(1989).} that they remain conserved when the theory is driven
off-criticality by the (thermal) $\Phi_{(1,3)}$ perturbation \KM\
\ref\EY{T. Eguchi and S.K. Yang, Phys. Lett. {\bf B224} (1989) 373;
T.J. Hollowood and P. Mansfield, Phys. Lett. {\bf B226} (1989) 73.}.

The first nine qKdV (or equivalently $\Phi_{(1,3)}$ perturbed minimal
models) conservation laws have been constructed and checked to commute
\ref\KATO{A. Kato,"{\it Integrable deformation of conformal field theory}",
Tokyo preprint (1991).}.
As was already clear from the first few, they have the same structure as
their classical counterparts. In particular there exists a unique
conservation law for each odd degree (=dimension or spin). The essential
difference, apart from normal ordering, is that the various coefficients
are now polynomials in the central charge $c$, with degree fixed by the
dimension of the conservation law.

In the study of the $\Phi_{(1,3)}$ perturbed $(p,p')=(2,2k+1)$
non unitary minimal models, it has been observed that the conservation laws
with dimension a multiple of $2k-1$ vanish: their density was found to be
exactly proportional to the singular vector of level $2k$ in the
vacuum representation (modulo a total derivative)
\ref\FKLME{P.G.O. Freund, T.R. Klassen and E. Melzer, Phys. Lett. {\bf B229}
(1989) 243.}.
This connection between vacuum singular vectors and qKdV conservation laws
is truly remarkable.
{}From the explicit expression for these vectors
\ref\BSA{L. Benoit and Y. Saint-Aubin, Phys. Lett. {\bf B215} (1988) 517;
M. Bauer, P. Di Francesco, C. Itzykson and J.B. Zuber to appear in Nucl.
Phys. {\bf B} (1991).},
we get an explicit expression for the qKdV conservation laws for each odd
degree, albeit for a different (and unique) value of $c$
in each case. That the density of this conservation law is somewhat trivial
(being a singular vector) does not blemish the very non-trivial fact that
this conservation law commutes with the qKdV Hamiltonian (before the
singular vector is modded out!). Therefore a proof of the above observation
boils down to a constructive proof of the existence of a conservation law
of dimension $2k-1$ for a value of $c$ outside the classical range
($c$ very large), namely for $c= 1 - 3(2k-1)^2/(2k+1)$.
This results in further support for the complete integrability of the
qKdV equation.

Such a proof has been given in
\ref\EYP{T. Eguchi and S.K. Yang, Phys. Lett. {\bf B235} (1990) 282.},
using the Feigin-Fuchs representation. Then, commutativity with the qKdV
Hamiltonian is equivalent to commutativity with the Sine-Gordon Hamiltonian
$\oint \sin \alpha_{-} \phi$ ($\phi$ a free field, $\phi(z) \phi(w) \sim
-\log(z-w)$, $\alpha_{-} = -(2p/p')^{1 \over 2}$ and $\alpha_{+}=
(2p'/p)^{1 \over 2}$). On the other hand, the vertex operators $e^{\pm i
\alpha_{+} \phi}$ have respective dimension $1$ and $(2p'/p)-1$.
The latter is integer only if $p=2$. In that case, the anticommutator of
$\oint e^{-i \alpha_{+} \phi}$ with $\oint e^{i \alpha_{+} \phi}$ produces,
by construction, a qKdV conservation law of dimension $p'-2=2k-1$,
which, also by construction, has density proportional to the vacuum singular
vector of dimension $2k$.

Another argument has been given in
\ref\KUNASU{A. Kuniba, T. Nakanishi and J. Suzuki, "Ferro and
antiferro-magnetization in RSOS models", Tokyo preprint 1990.}.
It explains the vanishing of some integrals in the $(2,p')$
minimal model as a result of an enhancement of symmetry. This is based on a
level-rank type duality for coset models, which translates into the
equivalence of particular $W_n$ models for different values of $n$.
Denoting a (in general non-unitary) minimal $W_n$ model, characterized
by two integers $(p,p')$, by $W_n^{(p,p')}$,this equivalence reads:
$W_n^{(n,p')} \equiv W_{p'-n}^{(p'-n,p')}$. (Notice that in both cases,
$p$ takes the smallest allowed value, which is the dual Coxeter number
of the corresponding $SL(p)$ algebra).
For the particular case $n=2$ ($W_2$=Virasoro), this shows that any
$(2,p')$ minimal model is equivalent to some $W_{p'-2}$ model. The
integrable hierarchy associated with the latter is the $SL(p'-2)$
generalized KdV equation
\ref\DS{V.G. Drinfeld and V.V. Sokolov, J. Sov. Math. {\bf 30} (1985) 1975.},
for which classical conservation laws exist
at each degree not a multiple of $p'-2$. It is natural to expect
that no additional conservation law arises in the quantum case.
Therefore the $W_2^{(2,2k+1)}$ conservation laws of dimension a multiple
of $2k-1$ should vanish. This being a purely quantum effect, the conservation
laws can vanish only if their density is a vacuum singular vector.

In this letter we present a third and more direct proof. It does not rely on
the
Feigin-Fuchs representation (and on the qKdV-Sine-Gordon relationship).
Also it does not require any a priori knowledge of symmetry enhancement.
In the latter respect it can thus be applied to models for which no
duality relation is known.
Our essential point is to show that the relation between vacuum singular
vectors
and qKdV conservation laws is encoded in a very simple way in the structure
of the first few submodules of the vacuum Verma module.
We then give the $N=1,2$ supersymmetric extensions of this result.
In the $N=1$ case, the related conservation laws are those of the
superconformal minimal models $(p=2,p'=4k)$, perturbed by the superfield
${\widehat \Phi}_{(1,3)}$. These are just the conservation laws of the
quantum version of the supersymmetric KdV equation
\ref\SUSY{P. Mathieu, Nucl. Phys. {\bf B336} (1990) 338.}.
For $N=2$, the vacuum singular vectors yield conservation laws of
the minimal models perturbed by the lowest dimension generator of
the chiral ring. These models are naturally associated to a $N=2$
supersymmetric KdV equation
\ref\FMVW{P. Fendley, S.D. Mathur, C. Vafa and N. Warner, Phys. Lett.
{\bf B243} (1990) 257; P. Fendley, W. Lerche, S. Mathur and N. Warner,
Nucl. Phys. {\bf B348} (1991) 66.}
\ref\SUSYII{P. Mathieu and M. Walton, Phys. Lett. {\bf B259} (1991) 106.}.
Finally, from the KNS duality \KUNASU, it is also clear
that a similar situation is
to be expected for all $W_{n}^{(n,p')}$ models, for $n \geq 3$. We show
explicitly in the $W_3$ case how the structure of the first vacuum submodules
directly implies a relation between singular vectors and conservation
laws of the quantum Boussinesq equation.

\newsec{Virasoro minimal models}

The qKdV equation is given by \KM:
\eqn\comm{ \partial_{t} T = [T,H] }
where
\eqn\ham{ H = \oint (TT) d\zeta }
and the parenthesis indicates normal ordering:
\eqn\normo{ (AB)(w) = \oint {{d\zeta} \over {\zeta-w}} A(\zeta) B(w). }
In terms of modes, one has
\ref\BBSS{F,.A. Bais, P. Bouwknegt, K. Schoutens and M. Surridge, Nucl.
Phys. {\bf B304}
(1988) 348.}:
\eqn\no{ (AB)(z)= \sum_{n=-\infty}^{\infty} (z-w)^{-n-h_A-h_B} (AB)_n(w)}
with
\eqn\nor{ (AB)_n= \sum_{m \leq -h_A} A_m B_{n-m} + \sum_{m > h_A}
B_{n-m} A_{m} }
$h_A$ and $h_B$ are the conformal dimensions of $A$ and $B$ respectively.
The qKdV conservation laws are integrals $H_{n-1}=\oint h_n d\zeta$ such
that $[H_{n-1},H]=0$. Hence a necessary and sufficient condition for $h_n$ to
be a conserved quantity is that:
\eqn\com{ \oint_{0} dw \oint_{w} dz (TT)(z) h_n(w) =0 }
Expanding $(TT)(z)$ in modes, one sees that this is equivalent to \KATO:
\eqn\cns{ \oint dw (TT)_{-3} h_n(w) =0}
In other words, $(TT)_{-3}h_n$ must be a total derivative, i.e of the form
$L_{-1}(...)$.
The explicit form of
$(TT)_{-3}$ in terms of Virasoro modes ($T(z)=\sum
z^{-n-2} L_n$) is:
\eqn\tt{ (TT)_{-3}=2 \sum_{m \geq 0} L_{-m-2} L_{m-1}}

We now proceed to prove the following result: for the minimal models
$(p=2,p'=2k+1)$, the vacuum singular vector at level $2k$, denoted by $v_{2k}$,
is a qKdV conserved density. This amounts to prove that
\eqn\ttv{ (TT)_{-3} v_{2k} \simeq 0}
where the symbol $\simeq$ means equality modulo total derivative.
Since the action of the positive Virasoro modes vanish on a singular vector,
this is equivalent to
\eqn\calc{ (2L_{-2}L_{-1} + 2 L_{-3} L_0)v_{2k} \simeq 2(2k-1)L_{-3}v_{2k}
\simeq 0 }
The key point of our argument is to notice that $v_{2k}$ is itself degenerate
at level $3$. This singular vector
$v_{2k+3}$ reads:
\eqn\sviii{ v_{2k+3}=(L_{-3} -{1 \over {k+1}}L_{-1}L_{-2} +{1 \over{
2(k+1)^2}} L_{-1}^3)v_{2k} }
Moreover $v_{2k+3}$ also belongs to the Verma module built on $L_{-1}{\bf I}$,
the level $1$ descendent of the vacuum. Therefore, after quotienting the
vacuum tower by its first descendent (i.e. setting $L_{-1}{\bf I}=0$),
we find that $v_{2k+3}=0$, which immediately results in:
\eqn\res{ L_{-3} v_{2k} \simeq 0,}
and completes the proof of the above statement.

It is very easy to convince oneself that this scenario can only apply to
minimal models with $p=2$. Indeed the first singular vectors in the vacuum
module of a generic $(p,p')$ model occur at levels $1$, $p'-1$ and $p'+2p-2$.
To see this, recall that the singular vectors
of the vacuum module arise at levels
\ref\FF{B.L. Feigin and D.B. Fuchs, Funct. Anal. Appl. {\bf 17} (1983) 241;
A. Rocha-Caridi, in "{\it Vertex operators in Mathematics and Physics}",
MSRI Publications {\bf 3} (Springer 1984) 451.}:
\eqn\levsing{ {{(2npp'+p \pm p')^2 -(p-p')^2} \over {4pp'}} \ \ \ ,\ \ \
n \in {\bf Z} }
To make contact with qKdV conservation laws, a gap of dimension at most $3$
is required between $v_{p'-1}$ and $v_{p'+2p-2}$. This is possible only for
$p=2$ (one has $p \geq 2$ in the Virasoro case).
Notice that we need only concentrate on these first singular vectors, as
$v_{p'-1}$ is the only singular vector of the vacuum module which is not
a descendent of $L_{-1}{\bf I}$.

\newsec{N=1 superconformal minimal models}

Introducing the super energy-momentum tensor:
\eqn\sue{ \hT(Z)={1 \over 2} G(z) + \theta T(z)}
where $Z$ stands for the doublet $(z,\theta)$, one can write the supersymmetric
qKdV equation as \SUSY($dZ=dz d\theta$ and $D=\theta \partial_z +
\partial_{\theta}$):
\eqna\skdv
$$\eqalignno{
\partial_{t} \hT &= [\hT,\hH]  &\skdv a\cr
\hH &= \oint dZ (\hT D \hT) &\skdv b\cr}$$
Normal ordering is defined by:
\eqn\sno{ (\hA \hB) (Z_2)= \oint dZ_2 {\theta_{12} \over Z_{12}} \hA(Z_1)
\hB(Z_2) }
with $\theta_{12}=\theta_1 - \theta_2$ and $Z_{12}= z_1 - z_2
-\theta_1 \theta_2$.
Modes of the superfields are introduced in the natural way as:
\eqn\smo{ \hA(Z_1)= \sum_{r \in {\bf Z}+h_{\hat A}} Z_{12}^{-r-h_{\hat A}}
\hA_r(Z_2) + \theta_{12} \sum_{n \in {\bf Z}+h_{\hat A}+{1 \over 2}}
Z_{12}^{-n-h_{\hat A}-{1 \over 2}} \hA_n(Z_2) }
Supersymmetric qKdV conservation laws $\hH_{n-{1 \over 2}} = \oint dZ \hh_n$
are then characterized by the condition
\eqn\scomm{ \oint dZ_2 \oint dZ_1 (\hT D \hT)(Z_1) \hh_n(Z_2) = 0}
By expanding $\hT D \hT$ in modes and using the super Cauchy theorem:
\eqn\scauchy{ \oint dZ_1 \theta_{12}^a Z_{12}^{-n-1} f(Z_1)
={1 \over {n!}} (\partial^n D^{1-a} f)(Z_2)\ \ \ , \ \ \ (a=0,1)}
one can rewrite \scomm\ as:
\eqn\scns{(\hT D \hT)_{-3} \hh_n = G_{-{1 \over 2}}(...) \simeq 0 }

Now for the superconformal minimal models $(p=2,p'=4k)$, the first three
singular vectors of the vacuum tower arise at level $1 \over 2$
(${\hv}_{1 \over 2} = G_{-{1 \over 2}} {\bf I}$), $2k-{1\over 2}$ and $2k+1$.
The levels are given by \levsing\ divided by 2
\ref\MRC{A. Meurman and A. Rocha-Caridi, Comm. Math. Phys. {\bf 107} (1986)
263.}.
Let us now prove that $\hv_{2k-{1 \over 2}}$ is a conserved density for
\skdv a. In that case, \scns\ reduces to
\eqn\scal{ (4k-2) L_{-3} \hv_{2k - {1 \over 2}} \simeq 0}
On the other hand the explicit expression for the singular vector $\hv_{2k+1}$
is:
\eqn\ssing{ \hv_{2k+1}=(G_{-{3 \over 2}} -{1 \over {2k}} L_{-1}
G_{-{1 \over 2}}) \hv_{2k-{1 \over 2}} }
Now we observe that $\hv_{2k+1}$ is a descendent of both $\hv_{2k-{1 \over 2}}$
and $G_{-{1 \over 2}} {\bf I}$. Therefore it vanishes in the quotient of the
vacuum module by its first singular vector. Applying $G_{-{3 \over 2}}$
to the r.h.s. of \ssing, we obtain finally:
\eqn\sfin{ {{k+1} \over k} L_{-3} \hv_{2k-{1 \over 2}} \simeq 0}
which is the desired result.

Here again, one can ask whether $p=2$ is the only case for which some
conservation
laws are given by singular vectors.
In fact the level difference between the second and third singular vectors
for generic $(p,p')$ models is $p-{1\over 2}$. Thus with $p=3$ we still get
a level difference smaller than $3$. The relevant vectors are $\hv_{p'-
{1 \over 2}}$ and $\hv_{p'+2}$, with
$$\hv_{p'+2}=(G_{-{5 \over 2}}
+a G_{-{3 \over 2}} L_{-1}+bG_{-{1 \over 2}} L_{-1}^2 + cG_{-{1 \over
2}}L_{-2}) \hv_{p'-{1 \over 2}} $$
for some values of $a$, $b$ and $c$. Hence $G_{-{5 \over 2}}
\hv_{p'-{1 \over 2}} \simeq 0$ but this does not imply \scal.
Therefore $p=2$ is indeed the only possibility.

\newsec{N=2 superconformal minimal models}

There are three integrable perturbations of the $N=2$ superconformal
minimal models with
\eqn\ssc{ c= 3(1 - {2 \over {k+2}}) }
corresponding to perturbations by the chiral fields
${\tilde \Phi}_l$ of dimension
$l/2(k+2)$, with $l=1,2$ and $k$ \FMVW \SUSYII.
Each of these cases can be associated to a $N=2$ supersymmetric qKdV
equation \SUSYII. For the $l=1$ perturbed theory, there are conservation laws
for all values of the degree not a multiple of $k+1$, in which case the
conserved density is proportionnal to a vacuum singular vector \FMVW.
The Hamiltonian of the corresponding $N=2$ supersymmetric qKdV equation is
\SUSYII:
\eqn\ssham{ \tH = \oint dZ [ (\tT(\tT \tT))+ {{c-3} \over 16}
(\tT [D^+,D^-] \tT) ] }
Here we have introduced the $N=2$ super energy-momentum tensor
\eqn\ssem{ \tT(Z)=J(z)+ {1 \over 2} \theta^- G^+(z) -{1 \over 2}
\theta^+ G^-(z) +\theta^+ \theta^- T(z) }
and $Z$ now stands for $(z,\theta^+,\theta^-)$, $dZ=dz d\theta^-
d\theta^+$, the superderivatives being $D^{\pm}= \partial_{\theta^{\mp}} +
\theta^{\pm} \partial_z$.
Normal ordering is now defined by:
\eqn\ssno{ (\tA \tB)(Z_2) = \oint dZ_1 {{\theta_{12}^+ \theta_{12}^-}
\over Z_{12} } \tA(Z_1) \tB(Z_2) }
with $Z_{12}= z_1 - z_2
-\theta_{1}^+ \theta_{2}^- +\theta_{1}^- \theta_{2}^+$.  Finally our
normalizations are fixed by the commutator: %
\eqn\ssnorm{ \{G_r^+,G_s^- \} = 2L_{r+s} +2(r-s)J_{r+s} +{c \over 3}
(r^2 - {1 \over 4})\delta_{r+s,0}.}

Proceeding as before we find that a
conserved density $\th_n$ of the $N=2$ super qKdV equation
$\partial_t \tT = [\tT,\tH]$ must satisfy
\eqn\sscns{ \big[ (\tT(\tT \tT)) +{{c-3} \over 16} (\tT[D^+,D^-] \tT)
\big]_{-3}^{\theta^+ \theta^-} \th_n \simeq 0}
Here the symbol $\simeq$ means equality modulo terms of the form
$G^{\pm}_{-{1 \over 2}}(...)$, and the superscript $\theta^+ \theta^-$
reminds that this mode refers to the $\theta^+ \theta^-$ component of
the superfield inside the brackets.

We now argue that \sscns\ is always satisfied by $h_{k+1}=\tv_{k+1}$,
the third
vacuum singular vector (the two first ones are $\tv_{1 \over 2}^{\pm}=
G_{-{1 \over 2}}^{\pm} {\bf I}$).
In that case, \sscns\ can be rewritten (after a fair amount of algebra):
\eqn\sscal{ - {3 \over 2} {{k+1} \over {k+2}} [ L_{-3} - 4(k+2)J_{-1}J_{-2}]
\tv_{k+1} \simeq 0}
On the other hand, $\tv_{k+1}$ is itself degenerate at level $2$
\ref\BFK{W. Boucher, D. Friedan and A. Kent, Phys. Lett. {\bf B172} (1986)
316; E.B. Kiritsis, Int. J. Mod. Phys. {\bf A3} (1988) 1871.},
and the singular vector reads:
\eqn\ssdesc{
\eqalign{\tv_{k+3}&=\big[ -2(k+1)(k+2)J_{-2} +2(k+2)L_{-1}J_{-1}
-{{k+2} \over 2}G_{-{1 \over 2}}^- G_{-{3 \over 2}}^+ \cr
&+{{k+2} \over 2}
G_{-{1 \over 2}}^+ G_{-{3 \over 2}}^- + G_{-{1 \over 2}}^-
G_{-{1 \over 2}}^+ L_{-1} - L_{-1}^2 \big] \tv_{k+1} \cr} }
Moreover, this vector is also a descendent of $G_{-{1 \over 2}}^{\pm}
{\bf I}$, therefore vanishes in the quotient of the vacuum module
by its first two singular vectors. Acting then on \ssdesc\ by $J_{-1}$
yields, modulo total superderivatives:
\eqn\ssfin{ {{k+2} \over 4} [ L_{-3} - 4(k+2) J_{-1}J_{-2}] \tv_{k+1}
\simeq 0}
This is equivalent to \sscal\ and completes our proof.

\newsec{${\bf W}_{\bf 3}$ minimal models}

The quantum version of the Boussinesq equation is naturally defined in terms
of the $W_3$ algebra by \KM:
\eqn\qbous{ \partial_t T =[T,H] \ \ \ \ \ \partial_t W =[W,H] }
where
\eqn\hbous{ H= \oint dz W }
As before $T$ denotes the energy-momentum tensor and $W$ is a spin $3$
conserved current. The conservation laws $H_{n-1} = \oint dz h_n$ are
characterized by the condition:
\eqn\combou{ W_{-2} h_n \simeq 0}
This condition is satisfied by the vacuum singular vector $v_{p'-2}$
of the $W_{3}^{(p=3,p'=3k \pm 1)}$ minimal models . This vector is
itself doubly degenerate at level $2$, and one of the singular descendents
reads:
\eqn\wdesc{v_{p'}= [ W_{-2} -{2 \over {p'-1}} L_{-1} W_{-1}] v_{p'-2} }
Like all singular vectors in the vacuum module except $v_{p'-2}$,
$v_{p'}$ is a descendent of a combination of $L_{-1} {\bf I}$ and
$W_{-1} {\bf I}$
\ref\FZ{ V.A. Fateev and A.B. Zamolodchikov, Nucl. Phys. {\bf B280[FS18]}
(1987) 644.}.
As $W_{-1}{\bf I} ={1 \over 2}[W_0,L_{-1}] {\bf I}
={1 \over 2}W_0 L_{-1} {\bf I}$, the vector $v_{p'}$ vanishes in the
quotient of
the vacuum module by $L_{-1}{\bf I}$. This means that $W_{-2}v_{p'-2}
\simeq 0$, and that $\oint v_{p'-2}$ is a conservation law for the
quantum Boussinesq equation \qbous.

\newsec{Conclusion}

We have presented a new and simple proof of the relation between a
vacuum singular vector and some conservation laws for qKdV type
equations or equivalently for a special class of off-critical
minimal models.
For the quantum systems considered, this implies the existence, for a
specific value of the central charge, of an integral which commutes with
the defining Hamiltonian, at every degree for which a classical
conservation law exists.

Going back to the general problem of finding the conservation laws for these
quantum equations, recall that these are expected to be polynomials of
$c$ with degree fixed by the dimension of the law. We just provided one
particular value of each of these polynomials, for a value of $c$
function of the dimension of the conservation law. This suggests that
the actual conservation laws around these values of $c$ (and therefore
for all values of $c$, due to polynomiality) might be some simple
deformation of the corresponding singular vector expressions.
Although quite appealing, this program admittedly lacks a systematic line
of attack.

%

\listrefs
\end